%% file: realtdynamics.tex
\documentclass[aps,prl,reprint,superscriptaddress]{revtex4-2}

\usepackage{graphicx}
\usepackage{dcolumn}
\usepackage{bm}

\usepackage{amsmath}
\usepackage{amssymb}
\usepackage[normalem]{ulem}
\usepackage{dsfont}
\usepackage{varioref}
\usepackage[export]{adjustbox}
\usepackage{color}
\usepackage{braket}

\usepackage[colorlinks = true,linkcolor = blue, urlcolor  = cyan, citecolor = cyan, anchorcolor = blue]{hyperref}

\begin{document}

\preprint{APS/123-QED}

\title{Quantum Wake Dynamics from Distinct Spectroscopic Perturbations}

\author{Umesh Kumar}
\affiliation{Materials Science and Technology Division, Oak Ridge National Laboratory, Oak Ridge, Tennessee 37831, USA}%

\author{Gonzalo Alvarez}
\affiliation{Computational Sciences and Engineering Division, Oak Ridge National Laboratory, Oak Ridge, TN, USA
}

\author{David Alan Tennant}
\affiliation{Department of Physics and Astronomy, University of Tennessee, Knoxville, Tennessee 37996, USA}%
\affiliation{Department of Materials Science and Engineering, University of Tennessee, Knoxville, Tennessee 37996, USA}

\author{Satoshi Okamoto}
\affiliation{Materials Science and Technology Division, Oak Ridge National Laboratory, Oak Ridge, Tennessee 37831, USA}

\date{\today}


\begin{abstract} Quantum wake dynamics in quantum magnets have recently been inferred from the dynamical spin structure factor, which probes only a restricted class of local perturbations. Here, we show that resonant inelastic x-ray scattering (RIXS) selection rules act as an operator filter on fractionalized excitations, producing distinct quantum wakes in the spin-$\frac{1}{2}$ Heisenberg antiferromagnetic chain. Using explicit real-time evolution of single-spin and spin-conserving bond correlators, we find that the conventional spin response propagates up to the maximum spinon velocity, $v_s=\frac{\pi}{2} J$, whereas the bond channels concentrate their spectral weight into a slower dominant wake with $v\simeq 0.92 J$, while weaker components remain bounded by the full spinon light cone.  The corresponding momentum- and frequency-resolved responses map onto experimentally accessible RIXS channels, demonstrating that different spectroscopic perturbations resolve complementary pathways of many-body propagation beyond the neutron-scattering spin structure factor. Their inelastic spectral weights further provide access to quantum Fisher information, while equal-time bond sum rules connect the same spectroscopic channels to the ground-state energy. Because the same correlators can be prepared and measured on quantum hardware, they also define direct, experimentally anchored benchmarks for quantum simulations, particularly in frustrated and higher-dimensional magnets where controlled classical real-time calculations become challenging. \end{abstract}

\maketitle


Real-time dynamics in quantum magnets provide a direct window into the propagation of correlations, entanglement, and many-body excitations in strongly interacting systems. Recent inelastic neutron scattering (INS) experiments have shown that quantum wake dynamics in spin chains can be interpreted in terms of femtosecond real-time evolution of spin correlations~\cite{Scheie2022}. Since neutrons couple directly to local magnetic moments, INS naturally probes the dynamical spin structure factor. This makes neutron scattering a powerful probe of spin dynamics, but it also restricts the form of the perturbation that can be applied to the system. More recently, these real-time dynamics have been simulated using superconducting qubits, highlighting the relevance of such an approach for quantum computing by providing direct access to real-time evolution~\cite{lee2026benchmarking, GillesBuchs2026}.

Resonant inelastic x-ray scattering (RIXS) offers a complementary route. Unlike neutrons, photons couple to the electronic degrees of freedom of a material through local charge, orbital, and spin-dependent fields~\cite{RevModPhys.83.705}. As a result, RIXS can access a broader set of correlation functions beyond the conventional dynamical spin structure factor~\cite{Kumar2022}. In this sense, different RIXS channels act as distinct local perturbations of the quantum magnet, constrained by resonance conditions and selection rules. This ability to vary the perturbing operator provides a unique opportunity to probe different aspects of real-time quantum dynamics.

The nature of the perturbation is especially important in low-dimensional magnets, where different operators can generate qualitatively distinct dynamical responses. For example, the infinite-temperature Heisenberg chain has been widely discussed in the context of Kardar-Parisi-Zhang (KPZ) dynamics, yet recent work on higher-moment response functions has shown deviations from conventional KPZ scaling~\cite{Rosenberg2024}, and extensions of KPZ physics to two dimensions continue to broaden the scope of nonequilibrium spin dynamics~\cite{Widmann2026}. These developments highlight that real-time dynamics are not only properties of the Hamiltonian, but also of the operator used to perturb and probe the system.

RIXS is particularly well suited to this perspective. Although the restricted momentum transfer in transition-metal $L$-edge RIXS can limit access to the Brillouin-zone boundary in two-dimensional cuprates, the zone center remains directly accessible, in contrast to practical limitations in neutron scattering near the zone center. Moreover, complementary approaches such as Cu $K$-edge RIXS~\cite{Ishii2025} and oxygen $K$-edge~\cite{FSDIR, KUMARNJP2018} in 1D cuprates  provide additional pathways to probe magnetic spectroscopy. Thus, RIXS does not simply reproduce the information contained in INS, but provides experimentally accessible perturbations and response channels that can reveal otherwise hidden aspects of ultrafast spin dynamics~\cite{Kumar2022,FSDIR, PhysRevLett.106.157205,  PhysRevLett.112.147401}.

Here, we investigate the real-time dynamics of the one-dimensional Heisenberg antiferromagnet generated by the distinct local perturbations associated with RIXS response functions. Using explicit time evolution, we show how these perturbations generate measurable dynamical correlations equivalent to those appearing in the RIXS scattering cross-section. Our results establish RIXS as a platform for probing femtosecond real-time dynamics beyond the dynamical spin structure factor measured in INS.

Finally, we place these results in the broader context of quantum simulation and control. Although recent quantum-computing studies have demonstrated spin transport in finite spin chains~\cite{Arnab2026}, current hardware has limitations in system size and accessible time scale, making classical simulations essential for long-distance dynamics. At the same time, advances in quantum control of correlated excitations, including Hubbard excitons~\cite{Baykusheva2026}, point toward new ways of preparing, perturbing, and measuring nonequilibrium dynamics of quantum matter. The RIXS-based framework developed here provides a spectroscopic bridge between these directions by connecting controlled local perturbations, real-time many-body dynamics, and experimentally measurable response functions.\\

\begin{figure}
    \centering
    \includegraphics[width=\linewidth]{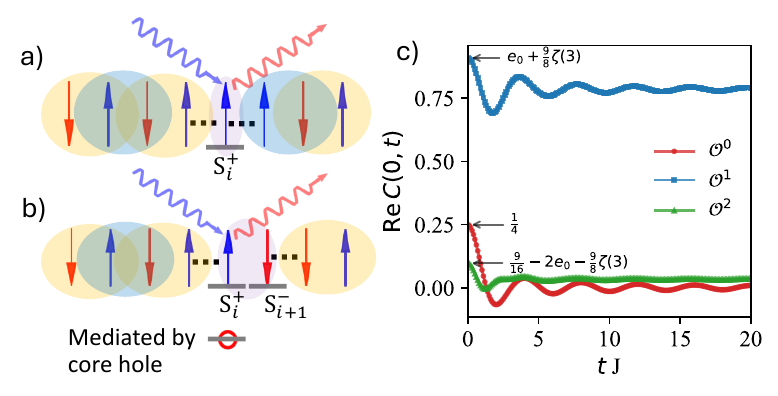}
\vskip -0.25cm
\caption{Schematics and local real-time dynamics of RIXS-relevant perturbations.
Panels (a) and (b) illustrate the single-spin and bond perturbations mediated by the core hole in the RIXS process. 
Panel (c) shows the real part of $C_\mathcal{O}(r=0, t)$ at the central site for all three perturbations: local single-spin, nearest-neighbor (NN) bond, and next-nearest-neighbor (NNN) bond operators.
}\label{fig:Schematics}
\end{figure}

{\sl Methods:---} We consider a spin-$\frac{1}{2}$ antiferromagnetic Heisenberg chain given by,
\begin{equation}\label{eq:hamiltonian}
H = J\sum_{\langle ij\rangle} {\bf S}_i\cdot{\bf S}_j ,
\end{equation}
where $J>0$ and $\langle ij\rangle$ denotes nearest-neighbor (NN) sites. RIXS at the Cu $L$-edge~\cite{PhysRevX.6.021020, Kumar2022, VivekKumar2026}, Cu $K$-edge~\cite{Ishii2025} and oxygen $K$-edge~\cite{FSDIR} involve distinct selection rules. Nevertheless, the RIXS response can be mapped to simpler correlation functions within ultra-short core-hole lifetime (UCL)~\cite{PhysRevB.75.115118,PhysRevX.6.021020, Kumar2022}. We therefore focus on the following three sets of operators: 
\begin{equation}\label{eq:Operators}
\begin{aligned}
a)~\mathcal{O}_i^0 &= S_i^\alpha, \quad b)~ \mathcal O_i^{1} = {\bf S}_i\cdot\left({\bf S}_{i+1}+{\bf S}_{i-1}\right),~\text{and}\\
c)~ \mathcal{O}_i^2 &= {\bf S}_{i-1}\cdot{\bf S}_{i+1}.
\end{aligned}
\end{equation}
Here, $\mathcal O_i^{0}$ represents a local-spin perturbation, while $\mathcal O_i^{1}$ and $\mathcal O_i^{2}$ represent spin-conserving bond perturbations corresponding to nearest neighbor (NN) and next-nearest-neighbor (NNN) bond operators, respectively~\cite{PhysRevB.77.134428, Forte2011, Kumar2022}. We use $\alpha=z$ for $\mathcal{O}_i^0$ in the calculations without loss of generality owing to spin-rotational symmetry for the local-spin perturbation.
The schematics for the local-spin and multi-spin perturbations are shown in panels (a) and (b) of Fig.~\ref{fig:Schematics}, respectively. These correlators are particularly relevant because they can be directly evaluated in quantum-computing simulations, as recently demonstrated for the local-spin response~\cite{Arnab2026}.

To connect real-time dynamics with RIXS-accessible response functions, we evaluate correlation functions of the form
\begin{equation}
C^\nu(r,t)
= \langle g| {\rm e}^{iHt} \mathcal O_{j}^\nu {\rm e}^{-iHt} \mathcal O_{i}^\nu |g\rangle ,  
\label{eq:real_time_corr}
\end{equation}
where $|g\rangle$ is the ground state wave function with the energy eigenvalue $E_g$, $r =R_j-R_i$, and $\mathcal O_i^\nu$ the perturbation operator and $\nu =\{0,1,2\}$ as defined in Eq.~\ref{eq:Operators}. 

To evaluate the correlators, we simulate Hamiltonian Eq.~(\ref{eq:hamiltonian}) with the density matrix renormalization group (DMRG)~\cite{PhysRevLett.69.2863, PhysRevB.48.10345}. We use DMRG++ \cite{re:alvarez0209}, two-site DMRG, and Krylov-space expansion of the Hamiltonian applied to the exponential~\cite{re:alvarez0713, PhysRevE.84.056706}; 
we have obtained converged results using $N=80$ sites, with $m = 250$ kept states; numerical parameters and details of the Krylov time evolution are given in the S3 of SM~\cite{SM}. 

The corresponding momentum and frequency-resolved response is obtained from the space-time Fourier transform
\begin{equation}
I^\nu(q,\omega)
=-\frac{2}{\pi}\int_0^{t_{\max}}dt\,e^{-\eta t}\sin(\omega t)
\sum_r\cos(qr)\,\mathrm{Im}\,C^\nu(r,t),
\label{eq:fourier_response}
\end{equation}
Implementation details are provided in S4 of the SM~\cite{SM}.  This quantity is directly analogous to the dynamical response measured in inelastic scattering experiments, with the important distinction that different choices of $\mathcal O_i^\nu$ correspond to different RIXS channels. 

\paragraph*{Quantum Fisher information.}
The same response functions provide access to multipartite entanglement through the quantum Fisher information~\cite{Hauke2016, https://doi.org/10.1002/qute.202400196, PhysRevLett.133.260202}. For an operator $\mathcal O_q = \frac{1}{\sqrt{N}} \sum_j {\rm e}^{iqR_j}\mathcal O_j$, the quantum Fisher information of the ground state is related to the equal-time fluctuation of $\mathcal O_q$. Equivalently, it can be obtained from the positive-frequency spectral weight
\begin{equation}
\mathcal F_Q(q)
= 4  \int_{0^+}^\infty \!\!  d\omega\,
I^\nu(q,\omega)/W_\mathcal{O},
\label{eq:QFI}
\end{equation}
Here, $W_{\mathcal O}$ is a normalization factor. $\mathcal F_Q(q)$ serves as a witness of multipartite entanglement; for the conventional local-spin generator, $\mathcal F_Q(q)>k-1$ implies at least $k$-partite entanglement. 
$W_{\mathcal O}
=(\lambda_{\max}-\lambda_{\min})^2,$
with ${\lambda}_{\text{max} (\text{min})}$ being the maximum (minimum) eigenvalue of ${\cal O}$.
For operators considered here, we have a) $W_{\mathcal{O}^0} =1$,  b) $W_{\mathcal{O}^1} =(\frac{3}{2})^2$ and c) $W_{\mathcal{O}^2} =1$~\cite{PhysRevB.103.224434, PhysRevB.111.205106, Cheng2022} (also see S2 in SM~\cite{SM}). 
Thus, the RIXS-accessible response functions provide a direct route to quantifying the entanglement encoded in the spin dynamics generated by different local perturbations.

 We further evaluate the  intensity integrated over the BZ and investigate the frequency dependence for the three cases to understand their similarities and differences.   \\



{\sl Results:---}   We first report the real-space and real-time correlation functions, in other words, the wake dynamics of operators appearing in RIXS spectroscopy evaluated for an $N=80$ chain. 

Fig.~\ref{fig:Schematics}(c) shows the real part of the local correlator $C_{\mathcal O}(r=0,t)$ at the center site. The red curve corresponds to the local-spin channel, $\mathcal O^{0}$ Its instantaneous value is fixed by the local spin sum rule, $C_{\mathcal O^{0}}(0,0)=\langle g|(S_c^\alpha)^2|g\rangle=\frac{1}{4}$.

The blue curve shows the NN bond response $\text{Re}[C_{\mathcal O^{(1)}}(0,t)]$, which remains positive over the time interval shown. In the thermodynamic limit, its equal-time value is fixed by the exact sum rule $W_1=C_{\mathcal O^{1}}(0,0)=e_0+\frac{9}{8}\zeta(3)\simeq0.909$~\cite{PhysRevLett.106.157205}, where $e_0=\frac{E_g}{NJ}=\frac14-\ln2\simeq-0.443$ is the ground-state energy per site in units of $J$~\cite{1938.Hulthen.AMAF.26A,PhysRevB.55.12510}, and $\zeta(3)$ denotes Apéry's constant. Similarly, the green curve represents the NNN bond response $\text{Re}[C_{\mathcal O^{2}}(0,t)]$, whose equal-time value satisfies $W_2=C_{\mathcal O^{2}}(0,0)=\frac{9}{16}-2e_0-\frac{9}{8}\zeta(3)\simeq0.096$, as derived in Sec.~S1 of the SM~\cite{SM}.

Importantly, combining the two spin-conserving channels yields the exact relation $W_1+W_2=\frac{9}{16}-e_0\approx 1.0$, providing direct access to the ground-state energy through
\begin{equation}
e_0=\frac{9}{16}-(W_1+W_2).
\end{equation}
Thus, the combined equal-time bond sum rule provides a direct route to determining the \textit{ground-state energy} of the Heisenberg chain.

\begin{figure}
    \centering
    \includegraphics[width=\linewidth]{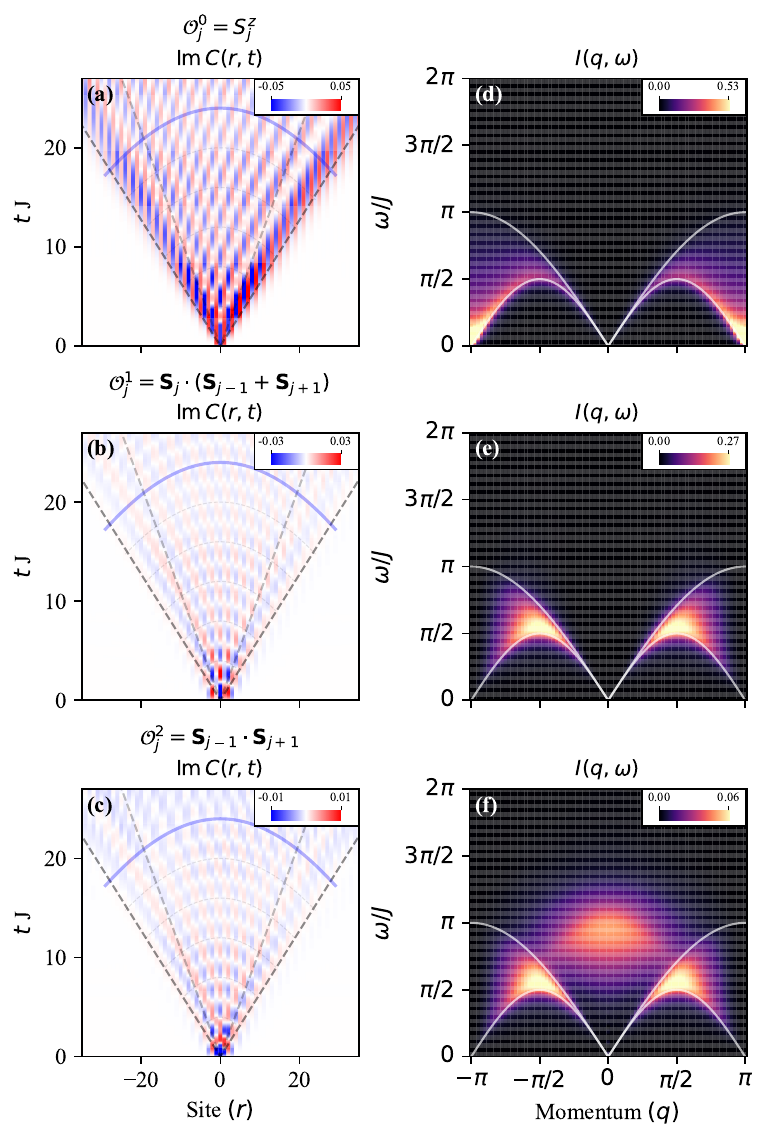}
\caption{Dynamics generated in the spin chain by RIXS-relevant perturbations. Panels (a)--(c) show the imaginary part of the space-time response for three perturbing operators: (a) the local-spin operator $S_i^\alpha$, (b) the nearest-neighbor (NN) bond operator $\mathcal{O}^1$, and (c) the next-nearest-neighbor (NNN) bond operator $\mathcal{O}^2$. The dashed lines indicate the outer spinon light cone with velocity $v_s=\frac{\pi}{2}J$ and the slower dominant wake contour with velocity $v_{\mathrm{dom}}=\frac{\pi \sqrt{3-\varphi}}{4}J$, corresponding to $t=\frac{r}{|v|}$, where $\varphi=\frac{1+\sqrt{5}}{2}$ is the golden ratio. The integrated $|\mathrm{Im}\,C(r,t)|$ in the intermediate region, $\frac{\pi \sqrt{3-\varphi}}{4}J<|v|<\frac{\pi}{2}J$, accounts for approximately $46\%$, $29\%$, and $28\%$ for the local-spin, NN-bond, and NNN-bond perturbations, respectively. Panels (d)-(f) show the corresponding momentum- and frequency-resolved responses.}
\label{fig:RealTimedynamics}
\end{figure}

Fig.~\ref{fig:RealTimedynamics} shows the real-time dynamics generated by the three perturbations. Panels (a)-(c) show the imaginary part of the spatially resolved correlation functions for the three operators. Panels (d)-(f) show the dynamical responses obtained by performing the double Fourier transform of the real-time correlation functions according to Eq.~\ref{eq:fourier_response}.

Panel (a) shows the response to a local-spin perturbation. In this case, the measured quantity is $\langle g|S_i^z|\Psi(t)\rangle$, which probes the projection of the time-evolved state onto the local spin orientation along the $z$ direction and the red and blue regions indicate positive and negative values of $\mathrm{Im}\,C(r,t)$, respectively. 
The disturbance propagates with a light-cone structure whose slope is set by the spinon velocity, $v_s=\frac{\pi}{2}J$, indicated by the black dashed lines. This behavior is consistent with previous real-time interpretations of spin-chain dynamics~\cite{Scheie2022}.

Panel (d) shows the conventional local-spin $I_{\mathcal{O}^{0}}(q,\omega)$  response. This response is directly accessible in INS~\cite{PhysRevLett.111.137205,Mourigal2013, Walters2009} and also constitutes the dominant magnetic contribution at the Cu $L_3$ edge of one-dimensional cuprates~\cite{Kumar2022}.

Panels (b) and (c) show the corresponding responses for the NN and NNN bond perturbations. Unlike the local-spin channel, these spin-conserving operators do not directly probe the local spin orientation. Instead, they measure the overlap $\langle g|\mathcal{O}_i|\Psi(t)\rangle$ and therefore track the propagation of bond operator correlations generated by the perturbation. 
For these channels, the dominant signal propagates with a reduced velocity, $v= \frac{\pi \sqrt{3-\varphi}}{4}J\approx 0.92 J$, where $\varphi=\frac{1+\sqrt{5}}{2}$ is the golden ratio. This dominant wake is most clearly resolved at short times. The choice of this velocity is justified in the discussion below.
A weaker response extends beyond this dominant cone, but remains bounded by the spinon light cone $v_s=\frac{\pi}{2}J$, as indicated by the black dashed lines.

Additionally, panel (c) exhibits a peculiar feature at the perturbation site: the correlator does not change sign over the initial short-time interval at the center site. This delayed sign reversal suggests a longer characteristic oscillation period for the NNN bond perturbation.

Quantitatively, within the wavefront indicated by the blue lines, the region $\frac{\pi \sqrt{3-\varphi}}{4}J<|v|<\frac{\pi}{2}J$ contains approximately $46\%$, $29\%$, and $28\%$ of the integrated $|\mathrm{Im}\,C(r,t)|$ for the local-spin, NN-bond, and NNN-bond responses, respectively. 

\begin{figure}
\centering
\includegraphics[width=1\linewidth]{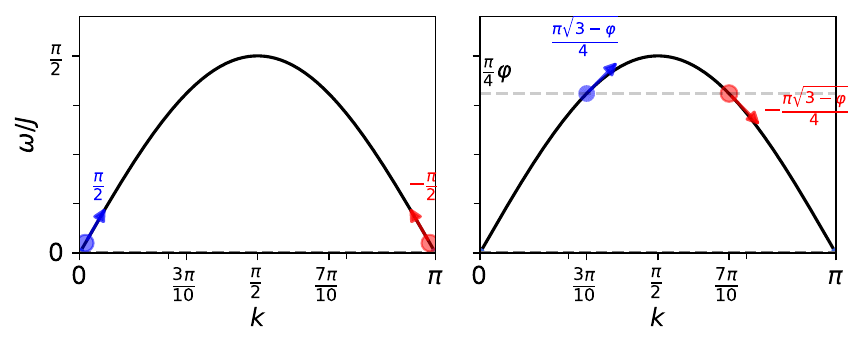}
\caption{Schematic illustration of the spinon dispersion and the momenta associated with the two dominant propagation velocities. The highlighted points indicate (a) the maximum spinon velocities, $v=\pm\frac{\pi}{2}J$, at $k={0,\pi}$ with $\omega=0$, and (b) the slower velocities, $v_k=\pm\frac{\pi \sqrt{3-\varphi}}{4}J$, at $k={\frac{3\pi}{10},\frac{7\pi}{10}}$ with $\omega = \frac{\pi}{4} \varphi J$.}
\label{fig:SchematicScattering}
\end{figure}

Panel (e) shows the dynamical response $I_{\mathcal{O}^{1}}(q,\omega)$ obtained from the Fourier transform of $C_{\mathcal{O}^{1}}(r,t)$. Importantly, the response vanishes at $q=\pm \pi$, in contrast to the local-spin case. The resulting spectrum reproduces the spin-conserving double-spin-flip response reported previously in the literature~\cite{Forte2011,PhysRevLett.106.157205, Ishii2025}. 

Panel (f) shows the dynamical response $I_{\mathcal{O}^{2}}(q,\omega)$ obtained from the Fourier transform of $C_{\mathcal{O}^{2}}(r,t)$. This channel captures the magnetic response probed by oxygen $K$-edge RIXS~\cite{FSDIR,Kumar2022,KUMARNJP2018} or the weaker signal in Cu $L_3$-edge~\cite{VivekKumar2026, Kumar2022} in one-dimensional cuprates. In contrast to the single-spin-flip response, its spectral weight vanishes at $q=\pm\pi$ and a distinct feature emerges at $q=0$.

These results establish the correspondence between the real-time dynamics and a range of experimentally relevant momentum-frequency correlators. They further demonstrate that distinct spectroscopic channels emphasize different regions of the Brillouin zone and provide complementary information about the underlying spin dynamics.

To motivate the propagation velocities identified above, we interpret the dynamics within a simple two-spinon picture, illustrated schematically in Fig.~\ref{fig:SchematicScattering}. The single-spinon dispersion of the Heisenberg chain is $\omega_s(k)=\frac{\pi}{2}J\sin k$ for $k\in[0,\pi)$~\cite{PhysRevB.52.13368,FADDEEV1981375,Kim2006}. Since the local-spin and bond operators considered here couple to states containing an even number of spinons, we focus on the lowest, two-spinon sector. For two spinons with momenta $k_1$ and $k_2$, the transferred momentum and excitation energy are $q=k_1+k_2\;(\mathrm{mod}\;2\pi)$ and $\Omega(k_1,k_2)=\omega_s(k_1)+\omega_s(k_2)$, respectively~\cite{Caux_2006}. Introducing the relative momentum $p=\frac{k_1-k_2}{2}$, such that $k_{1,2}=\frac{q}{2}\pm p$, the two-spinon center-of-mass velocity is
$v_{\mathrm{2sp}}(q,p)=\frac{\partial\Omega(q,p)}{\partial q}=\frac{\pi J}{2}\cos\left(\frac{q}{2}\right)\cos p$ (see Sec.~S5 of the SM~\cite{SM}).

\begin{figure}[t]
\centering
\includegraphics[width=\linewidth]{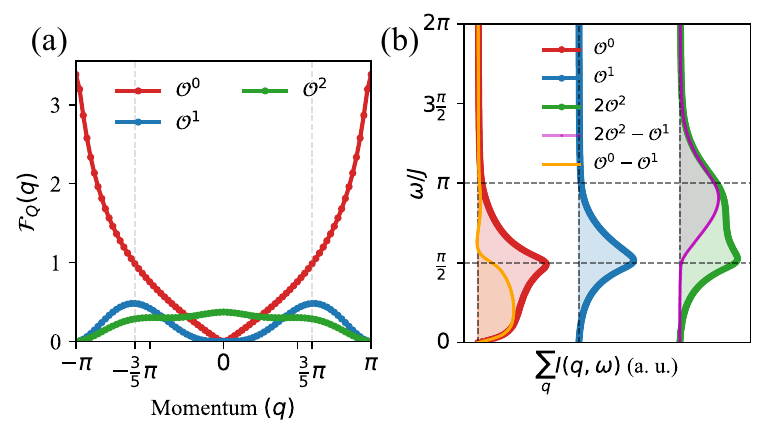}
\caption{Integrated spectral weights for the three responses. Panel (a) shows the quantum Fisher information extracted from the three RIXS-relevant correlation functions. Their distinct momentum dependence reflects the different quantum correlations accessed by the local-spin, nearest-neighbor bond, and next-nearest-neighbor bond perturbations. Panel (b) shows the corresponding momentum-integrated spectra over the Brillouin zone. The $\mathcal{O}^{2}$ response can be separated into a contribution resembling the $\mathcal{O}^{1}$ response and an additional feature appearing above $\frac{\pi}{2}J$.}\label{fig:QFIcomparison}
\end{figure}

The fastest propagation observed in the local-spin channel, $v=\pm\frac{\pi}{2}J$, is associated with the maximum individual-spinon velocity at the gapless momenta $k=0$ and $\pi$, as illustrated in Fig.~\ref{fig:SchematicScattering}(a). The two-spinon configurations $(k_1,k_2)=(0,\pi)$ and $(\pi,0)$ carry total momentum $q=\pi$, consistent with the dominant spectral weight near $q=\pi$ in Fig.~\ref{fig:RealTimedynamics}(d) and the corresponding maximum in the QFI shown in Fig.~\ref{fig:QFIcomparison}(a).

In contrast, the bond-operator responses concentrate their strongest spectral weight near $q=\pm\frac{3\pi}{5}$, as shown in Figs.~\ref{fig:RealTimedynamics}(e) and \ref{fig:RealTimedynamics}(f), and in the corresponding QFI in Fig.~\ref{fig:QFIcomparison}(a). For $q=\frac{3\pi}{5}$, the symmetric two-spinon configuration $p=0$ corresponds to $k_1=k_2=\frac{3\pi}{10}$, while the symmetry-related sector at $q=-\frac{3\pi}{5}\equiv\frac{7\pi}{5}$ corresponds to $k_1=k_2=\frac{7\pi}{10}$. These configurations are highlighted in blue and red, respectively, in Fig.~\ref{fig:SchematicScattering}(b). Each spinon has energy $\omega=\frac{\pi }{4}\varphi J$, and the two spinons propagate with the same velocity. The resulting center-of-mass velocity is therefore
\begin{equation}
	v_{\mathrm{dom}}=\frac{\pi J}{2}\cos\left(\frac{3\pi}{10}\right)
	=\frac{\pi \sqrt{3-\varphi}}{4} J
	\simeq0.923J.
\end{equation}
Thus, the slower dominant wake in the bond channels is naturally associated with the two-spinon configuration selected by the momentum region in which these operators concentrate their spectral weight.

Thus, the outer local-spin light cone reflects counterpropagating spinons at the maximum individual-spinon velocity, whereas the dominant bond wake is associated with the center-of-mass motion of a co-propagating two-spinon configuration selected near $q=\pm\frac{3\pi}{5}$.

We next examine the quantum Fisher information (QFI) associated with the three perturbations considered in this work. Fig.~\ref{fig:QFIcomparison}(a) shows the corresponding QFI obtained from Eq.~\ref{eq:QFI}. For the local-spin response, $\mathcal{F}_Q(q)$ vanishes at $q=0$ and reaches its maximum at $q=\pm\pi$. Under the standard normalization for local spin generators, the peak value witnesses at least four-partite entanglement, consistent with previous results obtained from INS~\cite{PhysRevB.107.059902,PhysRevB.103.224434}. In contrast, the QFI associated with $\mathcal{O}_i^{1}$ vanishes at both the zone center and zone boundary and is maximal near $q=\pm\frac{3\pi}{5}$, with $\mathcal{F}_Q(q)< 1$. The NNN response associated with $\mathcal{O}_i^{2}$, shown in green, instead peaks at $q=0$ and vanishes at $q=\pm\pi$. These contrasting momentum dependences demonstrate that the bond operators provide complementary, operator-resolved information about quantum correlations, including at the zone center where the conventional spin response vanishes.

Fig.~\ref{fig:QFIcomparison}(b) shows the spectra integrated over momentum, $\sum_q I(q,\omega)$, for the three responses. All three exhibit a pronounced maximum near $\omega=\frac{\pi}{2}J$, corresponding to the maximum of the lower boundary of the two-spinon continuum at $q=\frac{\pi}{2}$. A closer comparison shows that the NNN response contains two contributions: a dispersive component with an energy distribution similar to that of the NN response, and an additional higher-energy feature outside the dominant two-spinon continuum appears for $\omega>\frac{\pi}{2}J$. Conversely, subtracting the $\mathcal{O}^{1}$ response from the local-spin response isolates spectral weight for $\omega<\frac{\pi}{2}J$, the maximum energy of the single-spinon band. This further highlights an interesting relation between the momentum integrated response of these different correlators.

\paragraph*{Conclusions.—}
We have shown that quantum-wake dynamics in the Heisenberg chain depend strongly on the spectroscopic operator used to perturb and probe the system. The conventional local-spin channel produces a wake bounded by the maximum spinon velocity, $v_s=\frac{\pi}{2} J$. In contrast, the spin-conserving bond operators concentrate their spectral weight into a slower dominant wake with $v_\text{dom} = \frac{\pi\sqrt{3-\varphi}}{4} J \approx 0.92 J$, while a weaker component remains bounded by $v_s$. The distinct propagation velocities, local oscillation time scales, and momentum-resolved spectra demonstrate that RIXS selection rules select complementary regions of the spinon continuum. RIXS therefore encodes operator-dependent real-time dynamics beyond the conventional dynamical spin structure factor measured by INS.

The same experimentally accessible correlation functions provide information beyond excitation propagation. Their momentum-resolved spectral weights yield operator-dependent quantum Fisher information, and the equal-time sum rules for the spin-conserving channels connect the corresponding bond correlations to the ground-state energy. Importantly, these correlators also provide a direct route for benchmarking quantum-computing simulations against experiment: the RIXS-defined operators can be prepared, evolved, and measured on quantum hardware, and the resulting dynamics can be compared with the corresponding spectroscopic response. This connection becomes particularly valuable in two-dimensional, frustrated, and other strongly correlated systems where reliable classical simulations of long-time dynamics become prohibitively difficult. Our results thus establish experimentally measurable, operator-resolved quantum wakes as concrete targets for testing quantum simulators in regimes beyond the reach of conventional numerical methods.

{\sl Acknowledgments:---} This work was supported by the U.S. Department of Energy, Office of Science, Office of Advanced Scientific Computing Research and Office of Basic Energy Sciences, Scientific Discovery through Advanced Computing (SciDAC) program under the CON-NEQT project. D.A.T. is supported by the Quantum Science Center (QSC), a National Quantum Information Science Research Center of the U.S. Department of Energy (DOE).

\bibliography{references}

\clearpage
\onecolumngrid

\input{supplement}

\end{document}

%% file: supplement.tex

\setcounter{secnumdepth}{3}

\setcounter{section}{0}
\setcounter{equation}{0}
\setcounter{figure}{0}
\setcounter{table}{0}

\renewcommand{\thesection}{S\arabic{section}}
\renewcommand{\theequation}{S\arabic{equation}}
\renewcommand{\thefigure}{S\arabic{figure}}
\renewcommand{\thetable}{S\arabic{table}}

\begin{center}
	{\large\bfseries
		Supplemental Material for\\[0.2cm]
		``Quantum Wake Dynamics from Distinct Spectroscopic Perturbations''
	}
	
	\vspace{0.3cm}
	
	Umesh Kumar$^{1}$, Gonzalo Alvarez$^{2}$,
	David Alan Tennant$^{3,4}$, and Satoshi Okamoto$^{1}$
	
	\vspace{0.15cm}
	
	{\small
		$^{1}$Materials Science and Technology Division,
		Oak Ridge National Laboratory, Oak Ridge, Tennessee 37831, USA\\
		$^{2}$Computational Sciences and Engineering Division,
		Oak Ridge National Laboratory, Oak Ridge, Tennessee 37831, USA\\
		$^{3}$Department of Physics and Astronomy,
		University of Tennessee, Knoxville, Tennessee 37996, USA\\
		$^{4}$Department of Materials Science and Engineering,
		University of Tennessee, Knoxville, Tennessee 37996, USA
	}
\end{center}

\vspace{0.5cm}

\section{Relation between spin-conserving spectroscopic channels and the ground-state energy}

We establish the relation between the two spin-conserving spectroscopic responses and the ground-state energy, which allows us to derive a compact relation involving the corresponding bond operators. Recall that throughout this work we evaluate the real-time correlation function
\begin{equation}
C^\nu( r,t)
= \langle g| {\rm e}^{iHt} \mathcal O_{j}^\nu {\rm e}^{-iHt} \mathcal O_{i}^\nu |g\rangle .
\label{eq:real_time_corr}
\end{equation}
At equal time, $t=0$, and for operator positions,  $r=0$, $i.e.,~ i=j$, the local response reduces to
$C^\nu(r=0,t=0)=\langle g|(\mathcal O_i^\nu)^2|g\rangle$  and $\nu=\{1,2\}$ for the two bond operators considered in our work.
Thus, the instantaneous spectral response is directly determined by equal-time ground-state correlations. As shown below, for the two spin-conserving bond operators these correlations can be related to one another and, remarkably, to the ground-state energy of the Heisenberg chain.\\

For spin-$\frac12$, we have the relation,
\begin{equation}
S_i^\alpha S_i^\beta
= \frac14 \delta_{\alpha\beta}
+ \frac{i}{2}\epsilon_{\alpha\beta\gamma} S_i^\gamma ,
\label{eq:SiSi}
\end{equation}
which implies
\begin{equation}
(\mathbf S_i\!\cdot\!\mathbf S_j)^2
= \sum_{\alpha\beta} S_i^\alpha S_j^\alpha S_i^\beta S_j^\beta
= \sum_{\alpha\beta} S_i^\alpha S_i^\beta S_j^\alpha S_j^\beta
= \frac{3}{16} - \frac12\, \mathbf S_i\!\cdot\!\mathbf S_j .
\label{eq:sisdjsq}
\end{equation}

Using Eq.~\eqref{eq:SiSi}, one also finds
\begin{align}
(\mathbf S_i\!\cdot\!\mathbf S_j)(\mathbf S_i\!\cdot\!\mathbf S_k)
&= \sum_{\alpha\beta} S_i^\alpha S_j^\alpha S_i^\beta S_k^\beta \nonumber\\
&= \sum_{\alpha\beta} \left(\frac14\delta_{\alpha\beta}
+ \frac{i}{2}\epsilon_{\alpha\beta\gamma}S_i^\gamma\right)
S_j^\alpha S_k^\beta \nonumber\\
&= \frac14\,\mathbf S_j\!\cdot\!\mathbf S_k
+ \frac{i}{2}\,\mathbf S_i\!\cdot\!(\mathbf S_j\times \mathbf S_k) .
\label{eq:mixedidentity}
\end{align}

For $\mathcal{O}^1$, we evaluate $\langle g|(\mathcal O_i^1)^2|g\rangle$: 
\begin{equation}
\sum_i \left[\mathbf S_i\cdot(\mathbf S_{i+1}+\mathbf S_{i-1})\right]^2 ,
\end{equation}
expand
\begin{align}
\left[\mathbf S_i\cdot(\mathbf S_{i+1}+\mathbf S_{i-1})\right]^2
&= (\mathbf S_i\!\cdot\!\mathbf S_{i+1})^2
+ (\mathbf S_i\!\cdot\!\mathbf S_{i-1})^2 \nonumber\\
&\quad
+ (\mathbf S_i\!\cdot\!\mathbf S_{i+1})(\mathbf S_i\!\cdot\!\mathbf S_{i-1})
+ (\mathbf S_i\!\cdot\!\mathbf S_{i-1})(\mathbf S_i\!\cdot\!\mathbf S_{i+1}) .
\end{align}
Using Eqs.~\eqref{eq:sisdjsq} and \eqref{eq:mixedidentity},
\begin{align}
(\mathbf S_i\!\cdot\!\mathbf S_{i+1})^2
+ (\mathbf S_i\!\cdot\!\mathbf S_{i-1})^2
&= \frac38
-\frac12\left(\mathbf S_i\!\cdot\!\mathbf S_{i+1}
+\mathbf S_i\!\cdot\!\mathbf S_{i-1}\right),
\\
(\mathbf S_i\!\cdot\!\mathbf S_{i+1})(\mathbf S_i\!\cdot\!\mathbf S_{i-1})
+ (\mathbf S_i\!\cdot\!\mathbf S_{i-1})(\mathbf S_i\!\cdot\!\mathbf S_{i+1})
&= \frac12\,\mathbf S_{i+1}\!\cdot\!\mathbf S_{i-1},
\end{align}
since the chirality terms cancel in the symmetrized combination. Therefore
\begin{equation}
\left[\mathbf S_i\cdot(\mathbf S_{i+1}+\mathbf S_{i-1})\right]^2
= \frac38 -\frac12\left(\mathbf S_i\!\cdot\!\mathbf S_{i+1}
+\mathbf S_i\!\cdot\!\mathbf S_{i-1}\right)
+\frac12\,\mathbf S_{i+1}\!\cdot\!\mathbf S_{i-1}
\end{equation}
and hence, for periodic boundary conditions,
\begin{equation}\label{eq:NNflips}
\sum_i \left[\mathbf S_i\cdot(\mathbf S_{i+1}+\mathbf S_{i-1})\right]^2
= \frac{3N}{8} -\sum_i \mathbf S_i\!\cdot\!\mathbf S_{i+1}+\frac12\sum_i \mathbf S_i\!\cdot\!\mathbf S_{i+2}.
\end{equation} \\

For $\mathcal{O}^2$, we obtain $\langle g|(\mathcal O_i^2)^2|g\rangle$: 
\begin{equation}
\sum_i (\mathbf S_{i+1}\!\cdot\!\mathbf S_{i-1})^2,
\end{equation}
apply Eq.~\eqref{eq:sisdjsq} directly:
\begin{equation}
(\mathbf S_{i+1}\!\cdot\!\mathbf S_{i-1})^2
=
\frac{3}{16}
-\frac12\,\mathbf S_{i+1}\!\cdot\!\mathbf S_{i-1}.
\end{equation}
Thus
\begin{equation}
\sum_i (\mathbf S_{i+1}\!\cdot\!\mathbf S_{i-1})^2
= \frac{3N}{16} -\frac12\sum_i \mathbf S_{i+1}\!\cdot\!\mathbf S_{i-1}
\end{equation}
or equivalently, with periodic boundary conditions,
\begin{equation}\label{eq:NNNflips}
\sum_i (\mathbf S_{i+1}\!\cdot\!\mathbf S_{i-1})^2
=
\frac{3N}{16}
-\frac12\sum_i \mathbf S_i\!\cdot\!\mathbf S_{i+2}.
\end{equation}

Using Eq.~\ref{eq:NNflips} and \ref{eq:NNNflips}, 
\begin{equation}
\sum_i \mathbf S_i\!\cdot\!\mathbf S_{i+1} = \frac{9N}{16} - \sum_i \left[\mathbf S_i\cdot(\mathbf S_{i+1}+\mathbf S_{i-1})\right]^2
- \sum_i (\mathbf S_{i+1}\!\cdot\!\mathbf S_{i-1})^2 
\end{equation}

Therefore, using this relation one can evaluate the ground state energy
\begin{equation}
\frac{E_g }{J}= \frac{9N}{16} - \langle g| \sum_i \left[\mathbf S_i\cdot(\mathbf S_{i+1}+ \mathbf S_{i-1})\right]^2 |g\rangle
- \langle g|\sum_i (\mathbf S_{i+1}\!\cdot\!\mathbf S_{i-1})^2 |g\rangle
\end{equation}

We therefore obtain Eq.~(6) in the main text,
\begin{equation}  
e_0=\frac{E_g}{NJ}=\frac{9}{16}-(W_1+W_2). 
\end{equation} 
Here, $e_0$, is the ground state energy per site and $W_1 = \langle g|\frac{1}{N} \sum_i \left[\mathbf S_i\cdot(\mathbf S_{i+1}+ \mathbf S_{i-1})\right]^2 |g\rangle$ and $W_2 =\langle g|\frac{1}{N}\sum_i (\mathbf S_{i+1}\!\cdot\!\mathbf S_{i-1})^2 |g\rangle $. Using the exact result $W_1=e_0+\frac{9}{8}\zeta(3)$~\cite{PhysRevLett.106.157205}, Eq.~(17) immediately gives
$W_2=\frac{9}{16}-2e_0-\frac{9}{8}\zeta(3)$.
Thus, although the two individual spin-conserving sum rules contain Ap\'ery's constant $\zeta(3)$, these contributions cancel exactly in their sum, yielding Eq.~(17). 

\section{Eigenspectra of the operators}

We present the eigenspectrum needed for evaluating the quantum Fisher information reported in our work for the spin-$\frac{1}{2}$ chain~\cite{PhysRevB.103.224434}. We consider all three cases: i) $\mathcal O_c^{0}=S_c^z$, ii) $\mathcal O_c^{1}=\mathbf S_c\cdot(\mathbf S_{c-1}+\mathbf S_{c+1})$, and iii) $\mathcal O_c^{2}=\mathbf S_{c-1}\cdot\mathbf S_{c+1}$.\\

i) For the local operator,
\begin{equation}
\lambda=\frac{1}{2}:\quad \mathcal O_c^{0}\ket{\uparrow}=\frac12\ket{\uparrow},\qquad \lambda=-\frac12:\quad \mathcal O_c^{0}\ket{\downarrow}=-\frac12\ket{\downarrow}.
\end{equation}
The eigenvalue range is $\Delta\mathcal O=\lambda_{\max}-\lambda_{\min}=1$, giving the QFI normalization factor $W_{\mathcal O^0}=(\Delta\mathcal O)^2=1$.\\

ii) The operator $\mathcal O_c^1$ acts on three neighboring sites. We define $\mathbf T=\mathbf S_{c-1}+\mathbf S_{c+1}$ and $\mathbf J=\mathbf T+\mathbf S_c$. Since
\begin{equation}
\mathcal O_c^{1}=\frac12\left(\mathbf J^2-\mathbf T^2-\mathbf S_c^2\right),
\end{equation}
the sectors 
$(T,J)=(0,\frac{1}{2})$, $(1,\frac{1}{2})$, and $(1,\frac{3}{2})$ have eigenvalues $0$, $-1$, and $\frac{1}{2}$, respectively. Using the ordering $\ket{S_{c-1}^zS_c^zS_{c+1}^z}$, the eigenstates are 
\begin{align}
\lambda=-1:\quad &
\frac{\ket{\uparrow\uparrow\downarrow}-2\ket{\uparrow\downarrow\uparrow}+\ket{\downarrow\uparrow\uparrow}}{\sqrt6},\qquad
\frac{\ket{\uparrow\downarrow\downarrow}-2\ket{\downarrow\uparrow\downarrow}+\ket{\downarrow\downarrow\uparrow}}{\sqrt6},\\
\lambda=0:\quad &
\frac{\ket{\uparrow\uparrow\downarrow}-\ket{\downarrow\uparrow\uparrow}}{\sqrt2},\qquad
\frac{\ket{\uparrow\downarrow\downarrow}-\ket{\downarrow\downarrow\uparrow}}{\sqrt2},\\
\lambda=\frac12:\quad &
\ket{\uparrow\uparrow\uparrow},\qquad
\frac{\ket{\uparrow\uparrow\downarrow}+\ket{\uparrow\downarrow\uparrow}+\ket{\downarrow\uparrow\uparrow}}{\sqrt3},\\
&
\frac{\ket{\uparrow\downarrow\downarrow}+\ket{\downarrow\uparrow\downarrow}+\ket{\downarrow\downarrow\uparrow}}{\sqrt3},\qquad
\ket{\downarrow\downarrow\downarrow}.
\end{align}

We have the $W_{\mathcal O^1} = (\frac{3}{2})^2$ for the nearest-neighbor-bond operators. \\

iii) The operator $\mathcal O_c^{2}$ acts on the two sites $c-1$ and $c+1$. Using the ordering $\ket{S_{c-1}^zS_{c+1}^z}$,
\begin{align}
\lambda=-\frac34:\quad &\frac{\ket{\uparrow\downarrow}-\ket{\downarrow\uparrow}}{\sqrt2},\\
\lambda=\frac14:\quad &\ket{\uparrow\uparrow},\qquad \frac{\ket{\uparrow\downarrow}+\ket{\downarrow\uparrow}}{\sqrt2},\qquad \ket{\downarrow\downarrow}.
\end{align}
We have the $W_{\mathcal O^2} = 1$ for the next-nearest-neighbor bond operators considered in our work.

\section{DMRG++ implementation}
\subsection*{Krylov time evolution in DMRG++}

For \texttt{TimeStepTargeting} with \texttt{TSPAlgorithm="Krylov"}, DMRG++~\cite{re:alvarez0209} evaluates the real-time correlation function
\begin{equation}
C_{ij}(t)=\langle g|O_j^\dagger{\rm e}^{-{\rm i}(H-E_g)t}O_i|g\rangle=\langle g|O_j^\dagger|\phi_i(t)\rangle ,
\end{equation}
where
\begin{equation}
|\psi_i\rangle=O_i|g\rangle,\qquad H'=H-E_g,\qquad |\phi_i(t)\rangle={\rm e}^{-{\rm i}H't}|\psi_i\rangle .
\end{equation}
This procedure performs a real-time evolution~\cite{PhysRevE.84.056706}. 
The Lanczos recursion is initialized with the normalized source state  for $t=0^+$. 

\begin{equation}
|v_0\rangle=\frac{|\psi_i\rangle}{\beta_i},\qquad \beta_i=\sqrt{\langle\psi_i|\psi_i\rangle},
\end{equation}
and generates the $M$-dimensional Krylov basis
\begin{equation}
V_M=\left(|v_0\rangle,\ldots,|v_{M-1}\rangle\right),\qquad V_M^\dagger V_M=\mathbb I_M.
\end{equation}
The Hamiltonian projected onto this basis is the tridiagonal matrix
\begin{equation}
T_M=V_M^\dagger H'V_M.
\end{equation}
Since the source state is proportional to the first Lanczos vector, its Krylov-space representation is
\begin{equation}
V_M^\dagger|\psi_i\rangle=\beta_i\mathbf e_0,\qquad \mathbf e_0=(1,0,\ldots,0)^{\rm T}\in\mathbb C^M.
\end{equation}
The action of the propagator on the source state is approximated within the Krylov subspace with a systematically controlled truncation error. 
\begin{equation}
|\phi_i(t)\rangle\simeq V_M{\rm e}^{-{\rm i}T_Mt}V_M^\dagger|\psi_i\rangle=\beta_iV_M{\rm e}^{-{\rm i}T_Mt}\mathbf e_0.
\end{equation}

Diagonalizing the projected Hamiltonian as
\begin{equation}
T_M=U_MD_MU_M^\dagger,\qquad D_M={\rm diag}(d_0,\ldots,d_{M-1}),
\end{equation}
gives
\begin{equation}
|\phi_i(t)\rangle\simeq\beta_iV_MU_M{\rm e}^{-{\rm i}D_Mt}U_M^\dagger\mathbf e_0.
\end{equation}
Using $V_M\mathbf e_m=|v_m\rangle$ and $(U_M^\dagger\mathbf e_0)_\alpha=(U_M)_{0\alpha}^{*}$, the propagated state can be written explicitly as
\begin{equation}
|\phi_i(t)\rangle\simeq\beta_i\sum_{m=0}^{M-1}\sum_{\alpha=0}^{M-1}|v_m\rangle(U_M)_{m\alpha}{\rm e}^{-{\rm i}d_\alpha t}(U_M)_{0\alpha}^{*}.
\end{equation}
The correlation function follows by contracting the propagated state with the measurement state:
\begin{equation}
C_{ij}(t)\simeq\beta_i\sum_{m=0}^{M-1}\sum_{\alpha=0}^{M-1}\langle g|O_j^\dagger|v_m\rangle(U_M)_{m\alpha}{\rm e}^{-{\rm i}d_\alpha t}(U_M)_{0\alpha}^{*}.
\end{equation}

In time-step targeting, the evolution is divided into intervals of length $\tau$. Within each interval, DMRG++ constructs \texttt{TSPTimeSteps}$=n_v$ uniformly spaced target states,
\begin{equation}
|P_r(t)\rangle={\rm e}^{-{\rm i}H'\delta t_r}|\phi_i(t)\rangle,\qquad \delta t_r=\frac{r\tau}{n_v-1},\qquad r=0,\ldots,n_v-1,
\end{equation}
such that $|P_0(t)\rangle=|\phi_i(t)\rangle$ and $|P_{n_v-1}(t)\rangle=|\phi_i(t+\tau)\rangle$. All target states are generated from the same initial state using the Krylov representation of the effective Hamiltonian.

The DMRG basis is optimized to represent the ground state and the complete set of time targets through the weighted reduced density matrix
\begin{equation}
\rho_A=w_g\,{\rm Tr}_B|g\rangle\langle g|+\sum_{r=0}^{n_v-1}w_r\,{\rm Tr}_B|\widehat P_r(t)\rangle\langle\widehat P_r(t)|,\qquad w_g+\sum_r w_r=1,
\end{equation}
where $|\widehat P_r(t)\rangle$ denotes a normalized target state. After the prescribed number of DMRG sweep movements, the last target becomes the initial state of the next interval,
\begin{equation}
|\phi_i(t+\tau)\rangle\leftarrow|P_{n_v-1}(t)\rangle,
\end{equation}
and the Krylov representation and intermediate targets are reconstructed in the updated DMRG basis. This procedure is repeated until the required maximum time is reached.

With $\hbar=1$, an excitation of energy $\omega=E_n-E_g$ accumulates the phase $\omega t$ and has period $2\pi/\omega$. Consequently, $2\pi/J$ is the period only of a component with excitation energy $\omega=J$. \\

The calculations were performed for an open $N=80$ chain with $J=1$, $m=250$ kept states, $\Delta t=0.1/J$, $\tau=0.05/J$, $\texttt{TSPTimeSteps}=5$ (internal target spacing $0.0125/J$), adaptive Krylov dimension $M\leq 200$ with tolerance $10^{-12}$ and  $t_{\max}=27/J$.

\section{Mapping between real space, momentum, and frequency}
To make a direct connection with the spectroscopy, we transform the time-evolved operator to the  momentum and frequency space. 


We consider all three operators
\begin{equation}
\mathcal O_j^{0}=S_j^z,\qquad \mathcal O_j^{1}=\mathbf S_j\cdot(\mathbf S_{j-1}+\mathbf S_{j+1}),\qquad \mathcal O_j^{2}=\mathbf S_{j-1}\cdot\mathbf S_{j+1}.
\end{equation}
For each operator, DMRG++ evaluates the open-chain correlation function
\begin{equation}
C_c^\nu(r =j-c,t)=\langle g|\mathcal O_j^\nu{\rm e}^{-{\rm i}(H-E_g)t}\mathcal O_c^\nu|g\rangle ,
\end{equation}
where $c$ is a source near the center of the chain and $r_j=R_j-R_c$. For the bond operators, the sum over $j$ is restricted to the sites on which the corresponding operator is defined.

In a translationally invariant system, our spectral convention is
\begin{equation}
C^\nu(r,t)=\frac{1}{N}\sum_q{\rm e}^{-{\rm i}qr}\int_0^\infty d\omega\,{\rm e}^{-{\rm i}\omega t}I^\nu(q,\omega).
\end{equation}
Because the DMRG calculation uses a finite open chain and a single source, we approximate the momentum response using the even source-centered projection
\begin{equation}
C^\nu_{\mathrm I}(q,t)=\sum_j\cos[q(R_j-R_c)]\,{\rm Im}\,C_c^\nu(j,t).
\end{equation}
This projection gives $I^\nu(-q,\omega)=I^\nu(q,\omega)$ and reduces boundary asymmetries associated with using a single source.

Using the spectral representation of $C^\nu(q,t)$, the positive-frequency inelastic response can be recovered from its imaginary part as
\begin{equation}
I^\nu(q,\omega)=-\frac{2}{\pi}\int_0^{t_{\max}}dt\,{\rm e}^{-\eta t}\sin(\omega t)\,C^\nu_{\mathrm I}(q,t).
\end{equation}
The exponential factor ${\rm e}^{-\eta t}$ provides Lorentzian broadening. Since any time-independent elastic contribution to $C_c^\nu(j,t)$ is purely real, the sine transform of ${\rm Im}\,C_c^\nu(j,t)$ isolates the inelastic response without an explicit elastic-background subtraction.

For the uniform time grid $t_n=n\Delta t$, the integral is evaluated using the trapezoidal rule:
\begin{equation}
I^\nu(q_\ell,\omega_m)\simeq-\frac{2}{\pi}\sum_{n=0}^{N_t-1}w_n{\rm e}^{-\eta t_n}\sin(\omega_m t_n)\sum_j\cos[q_\ell(R_j-R_c)]\,{\rm Im}\,C_c^\nu(j,t_n),
\end{equation}
where $w_0=w_{N_t-1}=\Delta t/2$ and $w_n=\Delta t$ otherwise. We use $N+1$ display points
\begin{equation}
q_\ell=-\pi+\frac{2\pi\ell}{N},\qquad \ell=0,\ldots,N,
\end{equation}
with lattice spacing $a=1$. The points $q=-\pi$ and $q=\pi$ are physically equivalent and are both retained only to close the displayed Brillouin zone. In the reported spectra, we use $\eta=0.15J$, $t_{\max}=27/J$, and evaluate $\omega$ on 301 points over $0\leq\omega/J\leq2\pi$, subject to the Nyquist condition $\omega_{\max}<\pi/\Delta t$.




\section{Group velocity and real-space propagation in the two-spinon sector}

Consider two spinons with momenta $k_1$ and $k_2$ and single-spinon dispersion $\omega_s(k)=\frac{\pi J}{2}\sin k$ for $k\in[0,\pi)$. Introducing the total and relative momenta $q=k_1+k_2$ and $p=\frac{k_1-k_2}{2}$, we have $k_1=\frac{q}{2}+p$ and $k_2=\frac{q}{2}-p$. The corresponding two-spinon energy is
\begin{equation}
	\Omega(q,p)=\omega_s(k_1)+\omega_s(k_2)=\pi J\sin\left(\frac{q}{2}\right)\cos p.
\end{equation}

Differentiating with respect to the total and relative momenta gives
\begin{equation}
	v_q(q,p)=\frac{\partial\Omega}{\partial q}=\frac{1}{2}\left[v_s(k_1)+v_s(k_2)\right]=\frac{\pi J}{2}\cos\left(\frac{q}{2}\right)\cos p,
\end{equation}
and
\begin{equation}
	v_p(q,p)=\frac{\partial\Omega}{\partial p}=v_s(k_1)-v_s(k_2)=-\pi J\sin\left(\frac{q}{2}\right)\sin p,
\end{equation}
where $v_s(k)=\partial\omega_s(k)/\partial k=\frac{\pi J}{2}\cos k$. Here, $v_q$ describes the center-of-mass propagation of the two spinons, whereas $v_p$ describes their relative separation. The factor $1/2$ in $v_q$ follows from $\partial k_{1,2}/\partial q=1/2$, while no such factor appears in $v_p$ because $\partial k_1/\partial p=1$ and $\partial k_2/\partial p=-1$. The individual velocities are
\begin{equation}
	v_s(k_1)=v_q+\frac{v_p}{2},\qquad v_s(k_2)=v_q-\frac{v_p}{2}.
\end{equation}

Two distinct velocity scales appear in the real-space dynamics. The outer light cone is fixed by the maximum individual-spinon velocity, $v_s^{\mathrm{max}}=\frac{\pi J}{2}$. For the local-spin response, whose frequency-integrated spectral weight is strongest near $q=\pi$, a representative configuration is $(k_1,k_2)=(\pi,0)$. In this case, $q=\pi$ and $p=\frac{\pi}{2}$, giving
\begin{equation}
	v_q=0,\qquad v_p=-\pi J.
\end{equation}
Equivalently, the two spinons propagate with $v_s(\pi)=-\frac{\pi J}{2}$ and $v_s(0)=\frac{\pi J}{2}$. Thus, their center-of-mass velocity vanishes, while their relative motion generates the two leading fronts of the outer light cone.

The bond-operator response exhibits, in addition to this outer bound, a slower dominant wake. Its frequency-integrated spectral weight is maximal near $q=\pm\frac{3\pi}{5}$. At fixed $q$, the magnitude of the center-of-mass velocity $|v_q(q,p)|$ is maximal for $p=0$, corresponding to an equal-momentum, co-propagating pair with $k_1=k_2=q/2$.
The two spinons therefore propagate together with $v_s(k_1)=v_s(k_2)=v_q$. For $q=\frac{3\pi}{5}$, we have $k_1=k_2=\frac{3\pi}{10}$ and
\begin{equation}
	v_{\mathrm{dom}}=v_q=\frac{\pi J}{2}\cos\left(\frac{3\pi}{10}\right)=\frac{\pi J}{4}\sqrt{3-\varphi}\simeq0.923J,
\end{equation}
where $\varphi=\frac{1+\sqrt{5}}{2}$ is the golden ratio. The symmetry-related sector at $q=-\frac{3\pi}{5}$, equivalently $q=\frac{7\pi}{5}$, corresponds to $k_1=k_2=\frac{7\pi}{10}$ and gives $v_{\mathrm{dom}}\simeq-0.923J$. Thus, the maximum individual-spinon velocity determines the outer light cone, whereas the equal-momentum configurations selected by the bond operators produce the slower co-propagating two-spinon wake.

To quantify the weight between the dominant wake and the outer spinon light cone, we integrate $|\mathrm{Im}\,C^\nu(r,t)|$ over the region $v_{\mathrm{dom}}<|r/t|<v_s$ within the space-time window shown in Fig.~2 of the main text. Relative to the total integrated response, this region contains approximately $46\%$, $29\%$, and $28\%$ of the weight for the local-spin, NN-bond, and NNN-bond channels, respectively.

%